\documentclass[a4paper]{ltxdoc}

 \usepackage{graphicx}
 \usepackage{achemso}

\newcommand{\agt}{\lower2pt\hbox{$\:\stackrel{>}{
            \scriptstyle\sim}\:$}}
\newcommand{\alt}{\lower2pt\hbox{$\:\stackrel{<}{
            \scriptstyle\sim}\:$}}

\begin{document}

 \date{\today}

 \title{Electrical Conductance of Molecular Junctions by a Robust Statistical Analysis}

 \author{M.\ Teresa Gonz\'{a}lez\footnote{Email:\ \texttt{teresa.gonzalez@unibas.ch}\ \ \
  Web:\ \texttt{http://www.unibas.ch/phys-meso}}, Songmei Wu, Roman Huber, Sense J. van der Molen,\\Christian Sch\"{o}nenberger, and Michel Calame
  \\ Institut f\"ur Physik, Universit\"at Basel, Klingelbergstr.~82, CH-4056 Basel, Switzerland  }

 \maketitle


\begin{abstract}

We propose an objective and robust method to extract the electrical conductance of
single molecules connected to metal electrodes from a set of measured conductance
data. Our method roots in the physics of tunneling and is tested on octanedithiol
using mechanically controllable break junctions. The single molecule conductance
values can be deduced without the need for data selection.

\end{abstract}

To determine the feasibility of devices based on single molecules and to assess
their properties, a single or a few molecules have to be wired between two metal
electrodes. This has become a reality only recently through different techniques
such as scanning-probe microscopy, and mechanical and electromigration break
junctions~\cite{Cui2001,Donhauser2001,Xu2003,Agrait2003,Reed1997}. Using these
techniques, the electrical conductances $G$ of a broad range of molecular junctions
have been determined~\cite{Smit2002,Reichert2002,Salomon2003,McCreery2004} and
gating of single molecules has been
demonstrated~\cite{Park2002,Kubatkin2003,Champagne2005,Xu2005}.

These promising results are somewhat counterbalanced by the poor agreement in the
conductance values of single molecules reported by different groups. This
disagreement reflects our present poor insight into the atomistics of single
molecule junctions. To overcome junction-to-junction fluctuations, a statistical
analysis has been proposed, in which $G$ values of many junction realizations are
represented in a histogram. This analysis has first been implemented in atomic
junctions~\cite{Krans1993,Krans1995,Agrait2003}, and has subsequently been used in
metal-molecule junctions~\cite{Cui2001,Smit2002,Xu2003}. Peaks in the histogram
point to preferred junction geometries. Evidence for the formation of few-molecules
junctions was derived from the observation of a series of $G$ values appearing at
multiples of a fundamental single molecule value. The appearance of peaks in
$G$-histograms is a very striking observation. However, to resolve these, data
selection schemes have been applied~\cite{Xiao2004,Li2006,Haiss2004}. This situation
is unsatisfactory, because there is at present no generally accepted objective
selection criterion. We address this important question in this Letter.

As a test case, we have chosen octanedithiol
junctions.~\cite{Cui2001,Xu2003,Li2006,Suzuki2006,Haiss2004,Pobelov2006,Ulrich2006}
We use a mechanically controllable break junction
(MCBJ)~\cite{Moreland1985,Ruitenbeek1996} setup and acquire many conductance traces
in succession. We compare conductance histograms, which were generated with and
without data selection. We show that the conductance value assigned to a single
molecule is {\em robust}, and that data selections do not help to improve the
results. The most convincing representation is found in a histogram of $\log G$
rather than $G$.

\begin{figure}
 \includegraphics[width=8cm]{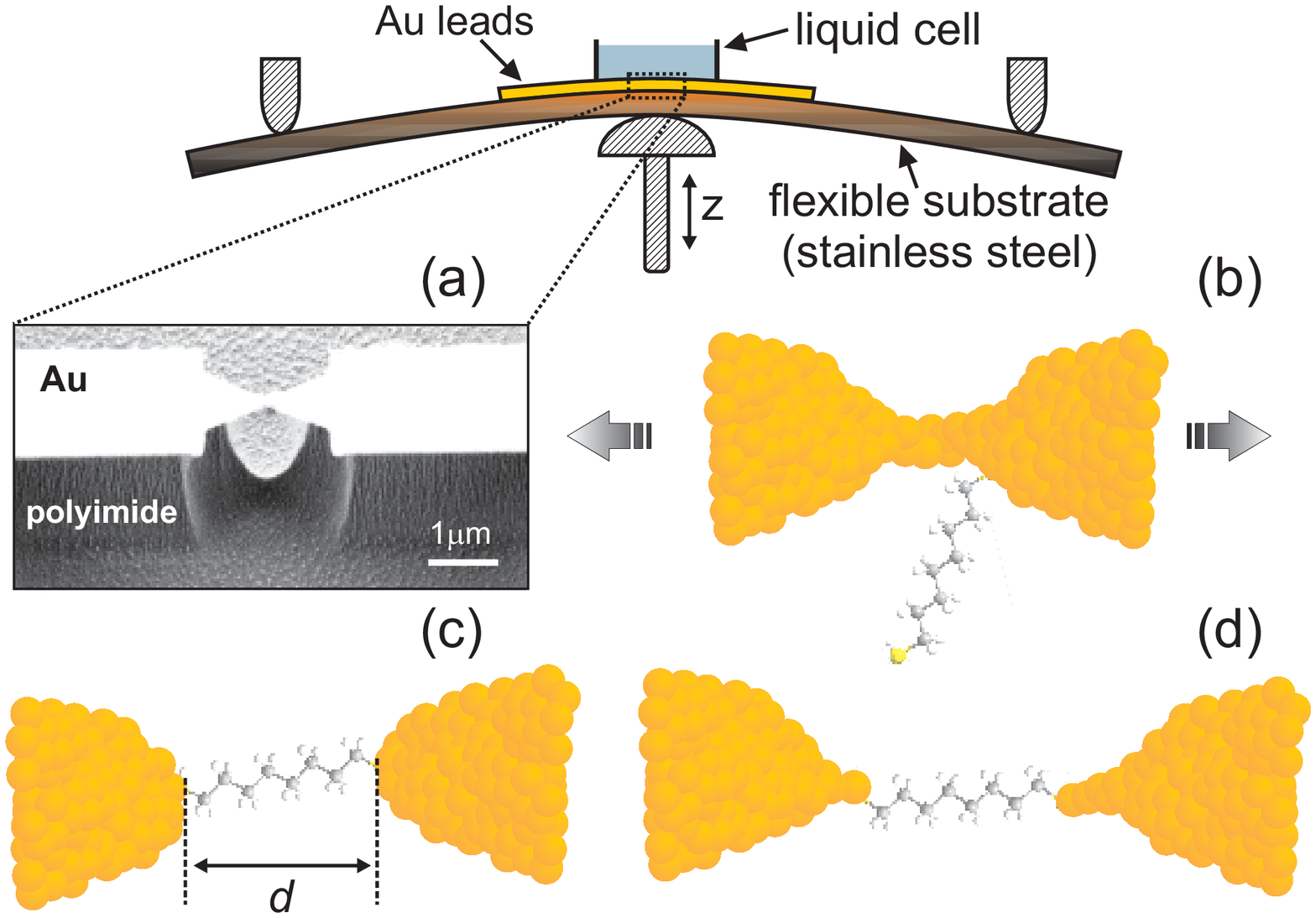}
\caption{\label{fig:Figure1} \footnotesize (a) Schematics of the MCBJ principle with
liquid cell and a SEM image of one underetched Au junction. (b)-(d) Principle of the
formation of a metal-molecule-metal bridge during the breaking process. Starting
from the fused Au leads (b), a molecule can lock between the leads (c). Under
further stretching, the Au leads are deformed, while the Au-octanedithiol-Au
junction stays intact (d).}
\end{figure}

The measurements were performed at room temperature, in a liquid environment.
Figure~\ref{fig:Figure1}a shows a schematics of our MCBJ setup~\cite{Grueter2005}.
As flexible substrates, we use electrically isolated stainless steel sheets, over
which gold leads are fabricated by electron beam lithography. The scanning-electron
microscope (SEM) image in Figure~\ref{fig:Figure1}a shows the suspended
\mbox{$100$\,nm}-wide region in the center of the gold leads. A flexible cell in the
top of the substrate guarantees that the leads are always immersed in liquid, (see
Figure~\ref{fig:Figure1}a). The substrate is held by two supports at the periphery
and a push-rod pressing from below. Bending the substrate results in stretching the
suspended Au bridge, which shrinks (b) until it breaks (c) and a gap of size $d$
forms. The change in gap-size $\Delta d$ is related to the vertical movement of the
push-rod $\Delta z$ by an attenuation factor $a=\Delta d/\Delta z = 1.6-4 \times
10^{-5}$~\cite{Grueter2005,Vrouwe2005}. The push-rod is moved at a velocity
\mbox{$v_z=30$\,$\mu$m/s}, so that the two Au leads separate at
\mbox{$0.5-1.2$\,nm/s}.

We apply a constant bias voltage of \mbox{0.2\,V}, and record the variation of the
current $I$ through the junction during repeated open-close cycles. The current was
measured with a current-voltage converter, which can automatically adjust its gain
between $0.1$ and \mbox{$100$\,V/$\mu$A}. This allows us to register the conductance
variation during the whole process, starting from the fused Au junction with
$G>G_0:=2e^2/h$, until the formation of single molecule junctions, with conductance
values orders of magnitude lower.

The formation of a metal-molecule bridge evolves in several steps. First, the
Au-bridge gets thinner (Figure~\ref{fig:Figure1}b), until a rather stable
single-atom contact is established. A plateau in the $G(z)$ curve is then expected
around $G_0$. When the atomic contact is finally lost, the conductance decreases
strongly. This decrease may be interrupted if a metal-molecule bridge is formed
(Figure~\ref{fig:Figure1}c). In that case, another plateau in the $G(z)$ curve is
anticipated.~\cite{Xu2003,Weber2002} Similar to atomic junctions, this
metal-molecule-metal bridge holds via its chemical bonds the two sides together and
postpones the breaking open of the Au electrodes (Figure~\ref{fig:Figure1}d).

\begin{figure}
 \includegraphics[width=8cm]{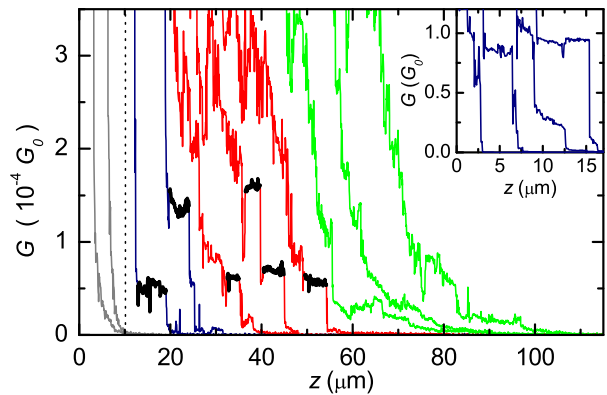}
\caption{\label{fig:Figure2} \footnotesize Variation of conductance during the
breaking process of a junction in pure mesitylene (left of the vertical pointed
line), and in a solution of octanedithiol in mesitylene (right). The curves are
shifted in $z$ for clarity. In the presence of octanedithiol, 50 \% of the curves
present plateaus. From these, some are very clean (the two first ones from left -
blue), and others are noisier (the following three ones - red). The remaining 50 \%
(the last three ones - green) show an irregular decay without plateaus. The plateaus
have been highlighted in black. Inset: Examples of plateaus close to 1 $G_0$,
corresponding to one-atom gold contacts.}
\end{figure}

To explore this process, we have performed groups of $100$ consecutive open-close
cycles for five different samples, both in pure mesitylene, and in a \mbox{1\,mM}
solution of octanedithiol in mesitylene. In Figure~\ref{fig:Figure2}, we show
representative $G(z)$ curves during opening of the bridge. Whereas the curves in the
main panel focus on values in the low conductance regime, i.e., at $G\approx
10^{-4}\,G_0$, the inset shows data around $G\approx G_0$, corresponding to the
single-gold-atom contact. Figure~\ref{fig:Figure2} shows how the shape of the
conductance curves $G(z)$ is modified by the presence of octanedithiol molecules.
Whereas $G(z)$ decays in an exponential fashion in the pure solvent (curves to the
left of the dotted line), distinct plateau features may appear in octanedithiol
containing solution. These plateaus are the signature of the formation of single
(few) molecule junctions. In some curves, jumps between plateaus at different $G$
can be seen. In those cases, the molecular junction reorganizes, and the number of
bridging molecules may change. In Figure~\ref{fig:Figure2}, the first two curves
(blue) to the right of the vertical dotted line, are rather ``clean''. In contrast,
the three next curves (red) are quite noisy just before plateau formation. This
suggests that there is a large degree of molecular movement in the junction, until
the octanedithiol molecules eventually lock between the leads. Finally, some $G(z)$
curves measured in the presence of molecules do not display plateaus (last three
curves in green). In this case, no stable single-molecule bridge has been formed.
Such traces correspond to approximately 50\% of the curves.

\begin{figure}
 \includegraphics[width=8cm]{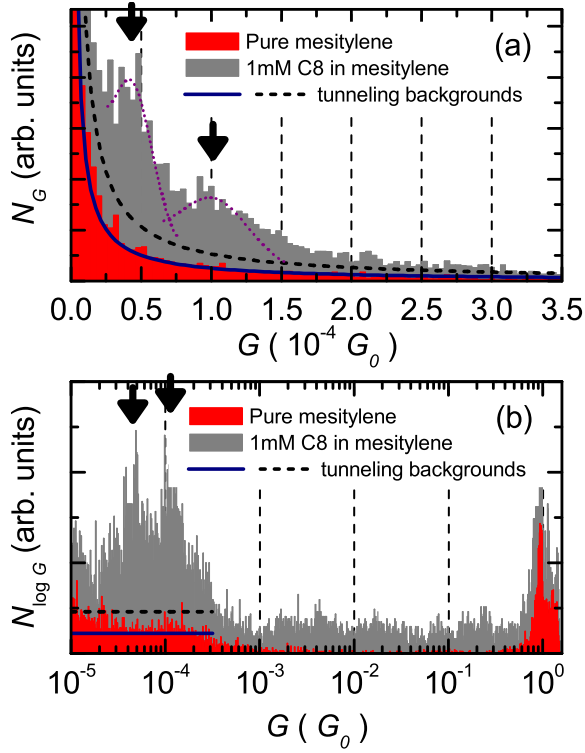}
\caption{\label{fig:Figure3} \footnotesize (a) Conductance histograms built from
approximately 100 $G(z)$ curves (Figure~\ref{fig:Figure2}) in pure mesitylene (red),
and in a solution 1mM of octanedithiol (grey). The arrows indicate the conductance
peaks that appear when octanedithiol is added in solution. The blue and black-dashed
lines show the best fit from below using a expression $\propto 1/G$ to both
histograms.(b) Histograms of $\log G$ built from the same data as in (a). The blue
and black-dashed lines correspond to the same $R/(2\,v_z \kappa a)$ values as in
(a). Note that in the latter representation both atomic gold peaks and molecular
peaks are observed.}
\end{figure}

Next, we focus on the statistics of our measured data. For each sample, we take all
$100$ conductance traces $G(z)$, and determine the probability with which a
particular $G$-value is measured, $N_G(G)$. This is depicted in the conductance
histograms of Figure~\ref{fig:Figure3}a (bin size: $\Delta G=4\times10^{-6}\,G_0$).
Whereas $N_G(G)$ decays smoothly in the pure solvent, there are distinct peaks
appearing in the octanedithiol case (indicated by arrows). This suggests that
particular molecular configurations form with a high probability. However, the peaks
in Figure~\ref{fig:Figure3}a are masked by a strong background. One may therefore
wonder, whether a particular data processing method could improve the sharpness of
the peaks. In the literature, different procedures have already been used, but they
have not carefully been compared with each other. In forming histograms, the
proposed procedures consist of (a) disregarding $G(z)$-curves that do not present
clear plateaus~\cite{Xiao2004,Xiao2005}, (b) only using data points that belong to
plateaus, instead of taking the whole $G(z)$ data~\cite{Pobelov2006}, (c) only using
average values derived from the data points belonging to conductance plateaus, and
weighting these by the plateau length~\cite{Li2006}; and d) using conductance
jumps~\cite{Haiss2004}. In focusing on the plateau values, these methods do
effectively eliminate a background. However, they can be subjective, as they involve
decisions as how constant the signal has to be to define a plateau, or where the
plateau exactly starts and ends.

We propose here an alternative method, which does not make use of any data
selection. We take all data, and only subtract a background that is adapted to the
physics of the problem. This method is as powerful as all the previous ones and,
most importantly, it is fully objective. In proceeding, we note that the conductance
must contain a tunneling contribution~\cite{Grueter2005}. The tunneling conductance
$G$ is exponentially dependent on the gap distance $d$, i.e., $G\propto \exp (-2
\kappa d)$. Here, $\kappa= \sqrt{2 m \phi}/\hbar$ is the decay constant, $\phi$ the
apparent barrier height, and $m$ the electron mass.  Furthermore, $d=a(z-z_0)$,
where $a$ is the attenuation factor of the MCBJ~\cite{Grueter2005,Vrouwe2005}, and
$z_0$ is defined as $z(d=0)$. Rewriting this, we find $ \ln G= -2\,\kappa\,a\,z +
constant$. It seems therefore much more appropriate to plot histograms of $\ln G$
rather than of $G$.

Making use of this expression, we can now calculate which is the expected tunneling
contribution in the conductance histograms. If we denote with $N_G$, $N_{\ln G}$,
and $N_z$ the respective probabilities of measuring a certain value of $G$, $\ln G$
and $z$, we may write
\begin{equation}\label{eq:equation1}
  N_G(G)\,dG = N_{\ln G}(\ln G)\,d\ln G = -N_z(z)\,dz.
\end{equation}
Here, $N_z(z)=R/v_z$, where $R$ is the data acquisition rate, and $v_z$ is the
velocity of the vertical push-rod. In our case, both these quantities are constant:
$R=500$~points/s, $v_z=30$~$\mu$m/s. Solving eq~\ref{eq:equation1} for $N_{\ln G}$
yields
\begin{equation}\label{eq:equation2}
  N_{\ln G}(\ln G)=\frac{R}{2\,v_z \kappa\, a}.
\end{equation}
Consequently, $N_{\ln G}$ is constant, whenever $\phi$ and $a$ are constants. Hence,
in a $\ln G$- or $\log G$-histogram, tunneling shows up as a constant background
which is easily subtracted. In Figure~\ref{fig:Figure3}b, we show a $\log
G$-histogram built from the data in Figure~\ref{fig:Figure3}a (bin size: $\Delta
\log(G/G_0) = 5\times10^{-3}$). A constant background is indeed present for $G\alt
2\times 10^{-4}\,G_0$ for the pure solvent (blue line), for which tunneling is the
only expected contribution. In contrast, clear peaks appear in the presence of
octanedithiol. The $\log G$-histogram representation is very powerful for another
reason: it presents a full overview of the data. At a glance, both the  single-atom
Au contact peaks ($G\approx G_0$) and the molecules signal ($G < 10^{-3}\,G_0$) are
seen. Between \mbox{$10^{-2}$--$10^{-3}\,G_0$} (depending on the sample) and $G_0$,
there is almost no weight in the histograms. This indicates that the Au atoms
retract quickly immediately after breaking the gold atom bridge.

The tunneling background, which is constant in a $\log G$-histogram, is inversely
proportional to $G$ in a $G$-histogram, the latter being the representation the
literature focused on so far. Solving eq~ \ref{eq:equation1} for $N_{G}$ yields
\begin{equation}\label{eq:equation3}
  N_{G}(G)=\frac{R}{2\,v_z \kappa a} \frac{1}{G}.
\end{equation}
As can be seen in Figure~\ref{fig:Figure3}a, this expression perfectly matches the
$G$-histogram of the pure solvent. The blue-line backgrounds of
Figure~\ref{fig:Figure3}a and b correspond to the same $R/(2\,v_z \kappa a)$.

\begin{figure}
 \includegraphics[width=8cm]{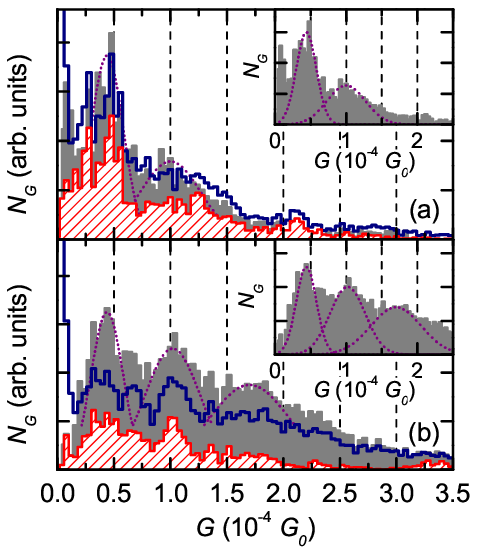}
\caption{\label{fig:Figure4} \footnotesize Histograms built from a given group of
$G(z)$ curves. (a) and (b) show results for two different samples. The grey
histograms were obtained considering all the measured conductance traces, and
subtracting later the tunneling background (also shown in the insets). The blue-line
histograms were made with the curves that display clear plateaus (blue and red
curves in Figure~\ref{fig:Figure2}). Finally, the red dashed histograms were built
only with the $G$ values which belong to plateaus (marked in thick black in
Figure~\ref{fig:Figure2}). The Gaussian curves highlight the position of the peaks.}
\end{figure}

We can use this property to subtract the tunneling background for the histograms on
dithiol molecules. To this end, we fit eq~\ref{eq:equation3} to our data from below.
This background is shown in Figure~\ref{fig:Figure3}a (black-dashed line). The same
$R/(2\,v_z \kappa a)$ gives the black-dashed constant background in
Figure~\ref{fig:Figure3}b. Subtracting it from the data yields a corrected histogram
which is guided by the physics of tunneling. The result of this subtraction is shown
in gray in Figure~\ref{fig:Figure4} (main panel and insets), for two different
samples. Figure~\ref{fig:Figure4}a corresponds to the data of
Figure~\ref{fig:Figure3}. From this analysis we conclude that junctions with
conductance values at multiples of $4.5\times10^{-5}\,G_0$ are more favorably
formed. This number is then assigned to the conductance of a single
Au-octanedithiol-Au bridge, $G_1$.

We will next compare our background subtraction method with other approaches based
on curve selection. This comparison is shown in Figure~\ref{fig:Figure4}. The
blue-line histogram has been obtained by taking only curves in which plateaus are
apparent (i.e. the blue and red curves in Figure~\ref{fig:Figure2}). In the red
dashed histogram only the points within a plateau have been used. The latter data
selection scheme is highlighted in black in Figure~\ref{fig:Figure2}. The selection
was done manually and no other treatment was applied.

On comparison of these three histograms, it is quite striking that all exhibit the
same key features. There are two, sometimes even three conductance peaks at
multiples of the same $G$-value (i.e., $G_1 \approx 4.5 \times 10^{-5}\,G_0$).
Particularly interesting is that the gray and blue-line histograms in
Figure~\ref{fig:Figure4} are almost identical. One can conclude from this that the
$G(z)$ curves without apparent plateaus can, on average, be described by a tunneling
dependence. The effective barrier height in this case is somewhat smaller than that
in the pure solvent. The third, red-dashed histogram, in which only plateau values
were considered also yields similar peak positions, but appears to have an even
stronger background subtracted. This is expected as in this histogram the noisy
signals away from the plateaus (as shown in the red curves of
Figure~\ref{fig:Figure2}) have been removed.

From the histograms, we find a single molecule conductance $G_1 \approx 4.5 \times
10^{-5}\,G_0$. In literature, values ranging from 1 to $25 \times 10^{-5}\,G_0$ have
been
reported.~\cite{Cui2001,Xu2003,Li2006,Suzuki2006,Haiss2004,Pobelov2006,Ulrich2006}
Our value lies very close to the one found by Wandlowski et al.\cite{Pobelov2006} It
is also close to that of Steigerwald et al.\cite{Venkataraman2006} for octanediamine
in trichlorobenzene ($2-6 \times 10^{-5}\,G_0$), which was obtained without the need
of any data selection. This similarity is particularly remarkable considering the
different bonding group of the molecules. Tao et al.~\cite{Li2006} reported two
groups of peaks, at multiples of $G_L=5.2 \times 10^{-5}\,G_0$ and multiples of
$G_H=2.5 \times 10^{-4}\,G_0$. They attributed these to two distinct microscopic
arrangements of the molecule-S-Au bonds. Whereas the first value agrees well with
our findings, we do not observe any other peak at higher conductance values. This is
especially made clear by the $\log G$-histogram of Figure~\ref{fig:Figure3}b. The
different solvent used in their work could be a possible cause for the formation of
the second group of peaks. However, Tao et al.~\cite{Li2006} observed peaks at the
same conductance values in different solvents. Another notable difference between
the two experiments is the speed at which the junctions are opened: \mbox{40\,nm/s}
in the work of Tao et al.~\cite{Li2006}, and \mbox{1\,nm/s} in our case. We
speculate that the change in speed could lead to the detection of different
microscopic conformations. Finally, a given microscopic arrangement could also be
favored in our symmetric MCBJ, in comparison with the more asymmetric junctions
formed in a scanning tunneling microscopy (STM) configuration.

From the above discussion, it is clear that a detailed analysis of conductance
histograms is required to gain insight in the microscopic formation of single
molecule junctions. In the analysis methods employed so far, a data selection
process has been used. In contrast, we demonstrate that a simple background
subtraction scheme suffices. It is as powerful as any data selection scheme and, in
contrast to the latter, it is objective. We emphasize that the statistical analysis
is most conveniently performed in a histogram in which $\log G$ is represented. The
$\log G$ representation allows a simple background subtraction and provides an
overview from the single atom contact to tunneling. Moreover, the single (few)
molecule conductance values show up in a much more striking manner. In addition, we
conclude that the features appearing in the conductance histograms obtained with
break junctions (in MCBJ or STM configuration) are robust and can be realistically
attributed to the molecular signature in these junctions.

We thank Th. Wandlowski for fruitful discussions and M. Steinacher for technical
support. S.J.vd.M. acknowledges the Netherlands Organisation for Scientific
Research, NWO ('Talent stipendium'), and M.T.G. the ``Ministerio de Educaci\'{o}n y
Ciencia'', for financial support. This work is supported by the Swiss National
Center of Competence in Research ``Nanoscale Science'', the Swiss National Science
Foundation, and the European Science Foundation through the Eurocore program on
Self-Organized Nanostructures (SONS).

{\bf Supporting Information Available:} Experimental details. This material is available free of charge via the Internet at
http://pubs.acs.org.

\newpage

\begin{thebibliography}{10}

\bibitem{Cui2001}
Cui,~X.~D.;\ \ Primak,~A.;\ \ Zarate,~X.;\ \ Tomfohr,~J.;\ \ Sankey,~O.~F.;\ \
  Moore,~A.~L.;\ \ Moore,~T.~A.;\ \ Gust,~D.;\ \ Harris,~G.;\ \ Lindsay,~S.~M.
  \textit{Science} \textbf{2001,} \textsl{294,} 571-574.

\bibitem{Donhauser2001}
Donhauser,~Z.~J.;\ \ Mantooth,~B.~A.;\ \ Kelly,~K.~F.;\ \ Bumm,~L.~A.;\ \
  Monnell,~J.~D.;\ \ Stapleton,~J.~J.;\ \ Price~Jr.,~D.~W.;\ \ Rawlett,~A.~M.;\ \
  Allara,~D.~L.;\ \ Tour,~J.~M.;\ \ Weiss,~P.~S. \textit{Science}
  \textbf{2001,} \textsl{292,} 2303-2307.

\bibitem{Xu2003}
Xu,~B.;\ \ Tao,~N.~J. \textit{Science} \textbf{2003,} \textsl{301,} 1221-1223.

\bibitem{Agrait2003}
Agra\"{i}t,~N.;\ \ Yeyati,~A.~L.;\ \ van Ruitenbeek,~J.~M. \textit{Physics
  Reports} \textbf{2003,} \textsl{377,} 81-279.

\bibitem{Reed1997}
Reed,~M.~A.;\ \ Zhou,~C.;\ \ Muller,~C.~J.;\ \ Burgin,~T.~P.;\ \ Tour,~J.~M.
  \textit{Science} \textbf{1997,} \textsl{278,} 252-254.

\bibitem{Smit2002}
Smit,~R. H.~M.;\ \ Noat,~Y.;\ \ Untiedt,~C.;\ \ Lang,~N.~D.;\ \ van
  Hemert,~M.~C.;\ \ van Ruitenbeek,~J.~M. \textit{Nature} \textbf{2002,}
  \textsl{419,} 906-909.

\bibitem{Reichert2002}
Reichert,~J.;\ \ Ochs,~R.;\ \ Beckmann,~D.;\ \ Weber,~H.~B.;\ \ Mayor,~M.;\ \
  v.~L\"{o}hneysen,~H. \textit{Phys. Rev. Lett.} \textbf{2002,} \textsl{88,}
  176804.

\bibitem{Salomon2003}
Salomon,~A.;\ \ Cahen,~D.;\ \ Lindsay,~S.;\ \ Tomfohr,~J.;\ \ Engelkes,~V.~B.;\
  \ Frisbie,~C.~D. \textit{Adv. Mater.} \textbf{2003,} \textsl{15,} 1881-1890.

\bibitem{McCreery2004}
McCreery,~R.~L. \textit{Chem. Mater.} \textbf{2004,} \textsl{16,} 4477-4496.

\bibitem{Park2002}
Park,~J.;\ \ Pasupathy,~A.~N.;\ \ Goldsmith,~J.~I.;\ \ Chang,~C.;\ \
  Yaish,~Y.;\ \ Petta,~J.~R.;\ \ Rinkoski,~M.;\ \ Sethna,~J.~P.;\ \
  Abru{\~{n}}a,~H.~D.;\ \ McEuen,~P.~L.;\ \ Ralph,~D.~C. \textit{Nature}
  \textbf{2002,} \textsl{417,} 722-725.

\bibitem{Kubatkin2003}
Kubatkin,~S.;\ \ Danilov,~A.;\ \ Hjort,~M.;\ \ Cornil,~J.;\ \
  Br\'{e}das,~J.-L.;\ \ Stuhr-Hansen,~N.;\ \ Hedeg{\aa}rd,~P.;\ \
  Bj{\o}rnholm,~T. \textit{Nature} \textbf{2003,} \textsl{425,} 698-701.

\bibitem{Champagne2005}
Champagne,~A.~R.;\ \ Pasupathy,~A.~N.;\ \ Ralph,~D.~C. \textit{Nano Lett.}
  \textbf{2005,} \textsl{5,} 305-308.

\bibitem{Xu2005}
Xu,~B.;\ \ Xiao,~X.;\ \ Yang,~X.;\ \ Zang,~L.;\ \ Tao,~N. \textit{J. Am. Chem.
  Soc.} \textbf{2005,} \textsl{127,} 2386-2387.

\bibitem{Krans1993}
Krans,~J.~M.;\ \ Muller,~C.;\ \ Yanson,~I.~K.;\ \ Govaert,~T. C.~M.;\ \
  Hesper,~R.;\ \ van Ruitenbeek,~J.~M. \textit{Phys. Rev. B} \textbf{1993,}
  \textsl{48,} 14721-14724.

\bibitem{Krans1995}
Krans,~J.~M.;\ \ van Ruitenbeek,~J.~M.;\ \ Fisun,~V.~V.;\ \ Yanson,~I.~K.;\ \
  de~Jongh,~L.~J. \textit{Nature} \textbf{1995,} \textsl{375,} 767-769.

\bibitem{Xiao2004}
Xiao,~X.;\ \ Xu,~B.;\ \ Tao,~N.~J. \textit{Nano Lett.} \textbf{2004,}
  \textsl{4,} 267-271.

\bibitem{Li2006}
Li,~X.;\ \ He,~J.;\ \ Hihath,~J.;\ \ Xu,~B.;\ \ Lindsay,~S.~M.;\ \ Tao,~N.
  \textit{J. Am. Chem. Soc.} \textbf{2006,} \textsl{128,} 2135-2141.

\bibitem{Haiss2004}
Haiss,~W.;\ \ Nichols,~R.~J.;\ \ van Zalinge,~H.;\ \ Higgins,~S.~J.;\ \
  Bethell,~D.;\ \ Schiffrin,~D.~J. \textit{Phys. Chem. Chem. Phys.}
  \textbf{2004,} \textsl{6,} 4330-4337.

\bibitem{Suzuki2006}
Suzuki,~M.;\ \ Fujii,~S.;\ \ Fujihira,~M. \textit{Jap. J. App. Phys.}
  \textbf{2006,} \textsl{45,} 2041-2044.

\bibitem{Pobelov2006}
Pobelov,~I.;\ \ Li,~Z.;\ \ Wandlowski,~T. to be pusblished.

\bibitem{Ulrich2006}
Ulrich,~J.;\ \ Esrail,~D.;\ \ Pontius,~W.;\ \ Venkataraman,~L.;\ \ Millar,~D.;\
  \ Doerrer,~L.~H. \textit{J. Phys. Chem. B} \textbf{2006,} \textsl{110,}
  2462-2466.

\bibitem{Moreland1985}
Moreland,~J.;\ \ Ekin,~J.~W. \textit{J. Appl. Phys.} \textbf{1985,}
  \textsl{58,} 3888-3895.

\bibitem{Ruitenbeek1996}
van Ruitenbeek,~J.~M.;\ \ Alvarez,~A.;\ \ Pi{\~n}eyro,~I.;\ \ Grahmann,~C.;\ \
  Joyez,~P.;\ \ Devoret,~M.~H.;\ \ Esteve,~D.;\ \ Urbina,~C. \textit{Rev. Sci.
  Instrum.} \textbf{1996,} \textsl{67,} 108-111.

\bibitem{Grueter2005}
Gr\"{u}ter,~L.;\ \ Gonz\'{a}lez,~M.~T.;\ \ Huber,~R.;\ \ Calame,~M.;\ \
  Sch\"{o}nenberger,~C. \textit{Small} \textbf{2005,} \textsl{1,} 1067-1070.

\bibitem{Vrouwe2005}
Vrouwe,~S. A.~G.;\ \ van~der Giessen,~E.;\ \ van~der Molen,~S.~J.;\ \
  Dulic,~D.;\ \ Trouwborst,~M.~L.;\ \ van Wees,~B.~J. \textit{Phys. Rev. B}
  \textbf{2005,} \textsl{71,} 035313.

\bibitem{Weber2002}
Weber,~H.~B.;\ \ Reichert,~J.;\ \ Weigend,~F.;\ \ Ochs,~R.;\ \ Beckmann,~D.;\ \
  Mayor,~M.;\ \ Alrichs,~R.;\ \ v.~L\"{o}hneysen,~H. \textit{Chemical Physics}
  \textbf{2002,} \textsl{281,} 113-125.

\bibitem{Xiao2005}
Xiao,~X.;\ \ Nagahara,~L.~A.;\ \ Rawlett,~A.~M.;\ \ Tao,~N. \textit{J. Am.
  Chem. Soc.} \textbf{2005,} \textsl{127,} 9235-9240.

\bibitem{Venkataraman2006}
Venkataraman,~L.;\ \ Klare,~J.~E.;\ \ Tam,~I.~W.;\ \ Nuckolls,~C.;\ \
  Hybertsen,~M.~S.;\ \ Steigerwald,~M.~L. \textit{Nano Lett.} \textbf{2006,}
  \textsl{6,} 458-462.

\end{thebibliography}
\end{document}